\documentclass{article}
\usepackage{epsfig}
\newcommand{\figeps}[2]{\centerline{\epsfig{file={#1},height={#2},clip=}}}
\begin{document}
\section*{EVOLUTION OF MIGRATING PLANET PAIRS IN RESONANCE\footnote{to be 
published in {\it Celestial Mechanics and Dynamical Astronomy}}}
\subsection*{S.Ferraz-Mello
\footnote{IAG-Universidade de S\~ao Paulo, Brasil {\rm (sylvio@usp.br)}},
C.Beaug\'e \footnote{Observat\'orio Astron\'omico, Universidad Nacional de 
Cordoba, Argentina} and {T.A.Michtchenko} \footnote{IAG-Universidade de S\~ao 
Paulo, Brasil}}
\date{}

\begin{abstract}
In this paper we present numerical simulations of the evolution of planets or 
massive satellites captured in the 2/1 and 3/1 resonances, under the action of 
an anti-dissipative tidal force. The evolution of resonant trapped bodies show 
a richness of solutions: librations around stationary symmetric solutions with 
aligned periapses ($\Delta \varpi = 0$) or anti-aligned periapses 
($\Delta \varpi = \pi$), librations around stationary asymmetric solutions in 
which the periapses configuration is fixed, but with $\Delta \varpi$ taking 
values in a wide range of angles. Many of these solutions exist for large 
values of the eccentricities and, during the semimajor axes drift, the 
solutions show turnabouts from one configuration to another. The presented 
results are valid for other non-conservative forces leading to adiabatic 
convergent migration and capture into one of these resonances.
\end{abstract}

\subsubsection* {Keywords: exoplanets, Galilean satellites, resonance, 
planetary migration, tidal forces.}


\section{Introduction}
The present paper considers the problem of a pair of planets (or planetary 
satellites) captured in the 2/1 and 3/1 mean-motion resonances, in which the 
inner mass is under the action of an anti-dissipative force that continuously 
increases its semimajor axis. Convergent migration of bodies and the action of 
non-conservative disturbing forces are subjects that have been explored by 
many authors (for a recent publication and bibliographical list, see Jancart 
2002). However, in almost all those studies, one of the bodies is assumed to 
have negligible mass and does not exert any gravitational action on the rest 
of the system. The scenario usually considered is thus the restricted 
three-body problem to which a non-conservative force is added.
In this work, we are interested in systems where the two planets 
(or satellites) have finite masses. As examples of such systems, we may cite 
some of the newly discovered extrasolar planetary systems (e.g. Gliese 876, 
HD82943, 55 Cnc) and, in our Solar System, some satellite pairs such as 
Io-Europa or Europa-Ganymede. Even if we expect a phenomenology showing many 
of the characteristics seen in the restricted models, only actual calculations 
can tell us how much of the dynamics is equivalent and, if not, which are the 
main changes.

The paradigm of a system of finite mass bodies captured into resonance is the 
system formed by the three inner Galilean satellites of Jupiter. If we denote 
by $\lambda_1$, $\lambda_2$ and $\lambda_3$, the mean longitude of Io, Europa 
and Ganymede, respectively, it was shown by Laplace that the motion of these 
bodies is such that the angle
\begin{displaymath}
\lambda_1-3 \lambda_2 + 2\lambda_3
\end{displaymath}
oscillates around $\pi$ with a small amplitude, $\sim 10^{-3}$ rad (Lieske, 
1998). Laplace also investigated the effect of a perturbation leading to an 
acceleration of the mean longitudes of one of the satellites. He showed that 
the mutual interaction of the three satellites would redistribute the effect 
among all the longitudes in such a way that the libration of the angle 
$\lambda_1-3 \lambda_2 + 2\lambda_3$ around $\pi$ would be conserved (see 
Tisserand 1896, Ferraz-Mello 1979). Nonetheless, the current tidal 
acceleration of the Galilean satellites is very small. Modern determinations 
of the rate acceleration/velocity, as deduced from old satellite eclipses and 
recent mutual events observations, are 
$(2.88\pm 0.10)\times 10^{-10} {\rm yr}^{-1}$ (Aksnes and Franklin 2000) and  
$(-0.074\pm 0.087)\times 10^{-10} {\rm yr}^{-1}$ (Lieske 1987). Heat 
dissipation observed on the surfaces of Io and Europa indicates that tides 
raised by Jupiter on the satellites are among the main dissipative agents at 
work in this system (see Yoder, 1979).

More recently, several resonant extrasolar planetary systems have been 
discovered with dynamical features suggesting that they have undergone a 
large-scale migration. In most cases, this orbital migration is believed to be 
due to an interaction of the planets with a residual planetesimals disk. In 
some extreme cases, tides have also been considered among the sources of 
orbital evolution; this seems to be the case of the putative planetary system 
around {HD 83443} (Wu and Goldreich, 2002) where the inner planet has the 
smallest semimajor axis among all currently known exoplanets. However, recent 
observations have not confirmed the existence of a pair of planets around 
that star. It is worth noting that many of the discovered exoplanets, 
including the inner planet of $\upsilon\,$And, orbit at distances smaller 
than 0.077 AU where tidal interaction with the central star should be 
important. All these planets are candidates to have an initial dynamical 
evolution similar to that shown by the Jupiter-Io system.
 
In this paper, we present a series of numerical simulations of the dynamical 
evolution of planetary satellites due to tidal effects, concentrating on the 
orbital variation after a capture in the 2/1 and 3/1 mean-motion resonances.
Except for a change in the timescale of the process, these results should also 
be representative of the evolution of extrasolar planets moving very close to 
the central star. 

\section{2-body tidal interaction}

The anti-dissipative force used in this paper corresponds to tides raised on 
the central mass $m_0$ (star or planet) by an orbiting body $m_1$ (planet or 
satellite). In addition, $m_1$ is interacting with a similar body $m_2$ placed 
in an orbit external to that of $m_1$. The model is planar and the spin axis 
of the central mass is assumed normal to the orbital plane. The two secondary 
bodies have finite masses but their sizes are neglected; they are thus 
considered as mass points. We will further assume that the tidal interaction 
only affects the inner body and not the outer one. This is, by far, our most 
stringent hypothesis. Tidal forces are inversely proportional to the 
$7^{\,\rm th}$ power of the distance to the central body; at the beginning of 
our simulations, we always considered ${a_1}/{a_2} < 0.6$,  making the tidal 
interaction of the outermost orbiting body $10^2$ times smaller, but in 
simulations where the bodies escape an early capture and ${a_1}/{a_2}$ grows 
significantly, the results of a simulation taking into account both tidal 
interactions may be very different.

Before presenting the physics of the used model, we should stress that the 
results presented in this paper were obtained in the frame of a more general 
investigation of the interplay of tides and resonance amongst finite mass 
bodies. For this reason, no averaging of the forces was used. Averaged 
equations are fine tools to assess the main effects in restricted there-body 
problems, but are less confident when a third massive body is interacting with 
the other two. The reason is simple: the averaged equations inside and outside 
resonances are not the same (see Ferraz-Mello, 1987). In a general problem, 
the semi-major axes have secular variations, resonance zones are crossed, one 
after another, and new phenomena are likely to occur which may not appear if 
the equations are, in some way, simplified. 

In view of this, the experiments presented in this work were performed via 
numerical simulations of the exact equations. Although more accurate, this 
approach leads to new problems. Since realistic tidal effects introduce 
extremely slow orbital variations, the integration times are prohibitive. 
Consequently, we have been obliged to enhance these perturbations to be able 
to obtain significant variations in the system, in reasonable integration 
times. In all our simulations, we increased the value of the ratio 
\begin{equation}
\chi = 2 \Delta t (\Omega - n)
\end{equation}
by some 2 or 3 orders of magnitude. Here, $\Delta t$ is the tide time lag, 
$\Omega$ is the rotation angular velocity of the extended body where the tide 
is raised and $n$ the orbital mean motion of the tidally interacting two-body 
system{\footnote {$\chi$ is roughly equal to the parameter 
$Q^{-1} \simeq \tan \varepsilon$ of the Kaula-Goldreich and MacDonald (1964) 
tide models. ($Q$ is the dissipation factor and $\varepsilon$ is the tide 
phase lag.)}}. The consequence is an acceleration of the evolutionary 
timescale by the same factor increasing $\chi$. However, this is not a real 
scaling of the problem because the factor $\chi$ only enters in one of the 
right-hand side terms of the equations of motion. It is important to keep in 
mind that the increase of $\chi$ must not be taken too large; the interplay 
of the migration due to the tide and one resonance depends on the speed with 
which the resonance is crossed, and the probability of capture at a crossed 
resonance lessens as $\chi$ grows. However, in this paper, we are more 
interested in the phenomena taking place after a capture, which are less 
affected by the enhancement of $\chi$ as long as this does not introduce 
large forced oscillations about stable stationary solutions.

\subsection{Tidal forces}

\begin{figure}[th]
\figeps{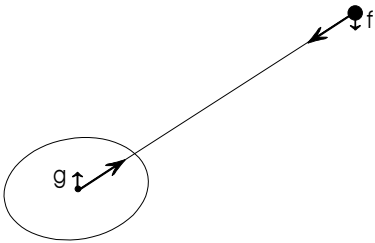}{35mm}
\caption{Interaction of an extended body and a mass point.${\bf f}$ and 
${\bf g}$ represent the resultant of the forces due to tides raised by the 
mass point on the extended body.}
\label{fig1}
\end{figure}

Consider a mass point $m_1$ raising tides on an extended body of mass $m_0$. 
Let ${\bf r}$ be the radius vector of $m_1$ relative to $m_0$. The big arrows 
in figure 1, show the central attraction forces between both masses, while
$\mathbf{f},\, \mathbf{g}$ are the resultants of the forces due to the tidally 
deformed shape of $m_0$. Since the system of forces is in equilibrium they 
satisfy to the relation ${\mathbf{f} = - \mathbf{g}}$. According with the 
Darwin-Mignard tidal model (Mignard 1981), we have:
\begin{equation}
{\mathbf{f}} = - {\displaystyle{3k_2\Delta t \frac{Gm_1^2 R_0^5}{r^{10}}}}
\left[2\mathbf{r} (\mathbf{r}\cdot \mathbf{v}) +
r^2(\mathbf{r}{\mathbf\times}\mathbf{\Omega} + \mathbf{v}) \right]
\label{eq1}\end{equation}
where $R_0$ and $\mathbf{\Omega}$ are the equatorial radius and the angular 
rotation velocity (vector) of the extended body, $k_2$ is the tidal Love 
number and $\mathbf{v}=\dot{\mathbf{r}}$. We can easily compute the 
acceleration of the two bodies with respect to an inertial frame and subtract 
them to obtain the relative acceleration:
\begin{equation}
\ddot{\mathbf{r}} = - \frac{G(m_0 + m_1) \mathbf{r}}{r^3} +
\frac{(m_0 + m_1){\mathbf{f}}}{m_0 m_1}
\end{equation}
or
\begin{equation}
\ddot{\mathbf{r}} = - \frac{GM \mathbf{r}}{r^3} + \frac{\mathbf{f}}{m}
\label{eq2}\end{equation}
where $M = m_0 + m_1$ is the total mass and $m$ is the reduced mass 
$m = \frac{m_0}{M} m_1$. If we assume that $m_0 \gg m_1$, then $m \simeq m_1$ 
and $M \simeq m_0$.

\subsection{The Torque on the Planet}
The momentum of the couple shown in figure \ref{fig1} is 
${\mathbf{r}{\mathbf\times} \mathbf{f}}$. 
Since the system is isolated, this momentum must be compensated by the 
momentum of the forces acting internally on the extended body with respect to 
its center of mass:
\begin{equation}
{\mathbf{T} = -\mathbf{r}{\mathbf\times} \mathbf{f}}.
\end{equation}
The rotation angular momentum $\mathbf{S}$ of the extended body will then 
change following the law
\begin{equation}
{\dot{\mathbf{S}} = \mathbf{T}}
\end{equation}
or, if we assume that the momentum of inertia $I$ of the extended body about 
the rotation axis is constant,
\begin{equation}
I\dot{\mathbf{\Omega}} = \mathbf{T}
\label{eq8}
\end{equation}
Introducing the expression for the force from equation (\ref{eq1}), we obtain
\begin{equation}
{\mathbf{T}} = {\displaystyle{3k_2\Delta t \frac{Gm_1^2 R_0^5}{r^{8}}}}
\left[r^2 \mathbf{\Omega} - (\mathbf{r}\cdot\mathbf{\Omega})\mathbf{v} +
\mathbf{v}{\mathbf\times} \mathbf{r} \right] .
\label{eqT1}
\end{equation}
Equations (\ref{eq8})-(\ref{eqT1}) give the temporal variation of the angular 
rotation velocity of the planet due to the tidal effects. In a series of 
preliminary simulations performed for the Jupiter-Io pair, we found that the 
perturbation is extremely small and the value of $\Omega$ remains practically 
unchanged. As an example, a typical run spanning 30 scaled Myrs (corresponding 
to 1.25 Gyr, considering the scaling of $\chi$) lead to a variation of the 
total orbital angular momentum of 
$\Delta S = +7.4 \times 10^{-11} M_{\,\odot} {\,\rm AU}^2/yr$ and, 
consequently, a variation of the rotation period of Jupiter of only +8.8 
seconds.

Following this result, for the rest of our calculations we will consider the 
rotation period of the extended body constant. Of course, a working hypothesis 
of this kind can only be adopted when the relative masses of the orbiting 
bodies to the central one are $< 10^{-3}$, as in the Jupiter-Io and many 
planet-exoplanet cases. In cases such as the Earth-Moon pair (MacDonald 1964, 
Touma and Wisdom 1994), the mass ratio is much higher ($\sim 1/80$) and the 
rotation period variations should be computed together with those of the 
orbital parameters.

\begin{figure}[hb!]
\figeps{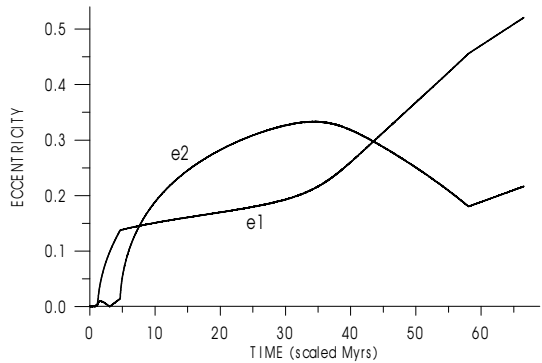}{5cm}
\caption{Eccentricities evolution after capture in the 2/1 resonance}
\label{fig5}
\end{figure}

\section{Evolution under Capture}

We now introduce our third body $m_2$ in the system, with an initial orbit 
exterior to $m_1$. We assume that $m_2$ does not have a tidal interaction with 
$m_0$ and that its orbit around $m_0$ is only perturbed by the gravitational 
interaction with $m_1$. When the semimajor axis of $m_1$ increases, 
mean-motion resonances between both satellites are reached. Capture then can 
take place. We performed a series of numerical simulations of the trapping 
process (and posterior orbital evolution within the resonance). The masses and 
initial conditions used in the simulation shown in this section were given by 
the following parameters:
\begin{displaymath}
\begin{array}{l@{\hspace{1cm}}l}
m_1 = 4.684 \times 10^{-5} m_0 &  m_2 = 2.523 \times 10^{-5} m_0\\
a_{1(0)} = 5.76 R_{0}	& 	a_{2(0)} = 9.40 R_{0}\\
e_{1(0)} = 0.0008 &	e_{2(0)} = 0.0003 \\
R_{0} = 7.14 \times 10^{7} m
& \Omega = 1.75 \times 10^{-4} s^{-1}\\
k_2\Delta t = 3 s\\
G=6.67 \times 10^{-11} m^{3}kg^{-1}s^{-2}.\\
\end{array}\end{displaymath}
The masses correspond to those of Io and Europa in units of Jupiter's mass.
$R_0$ and $\Omega$ are the equatorial radius and the angular rotation velocity 
of Jupiter and $k_2\Delta t$ is a scaled value, approximately 400 times the 
actual value for Jupiter (the estimated $Q$ of Jupiter is very large, see 
Yoder 1979, Peale 1999). The initial semimajor axis of the outer mass, 
$a_{2(0)}$, corresponds to the current semimajor axis of Europa; that of 
$a_{1(0)}$ is a little less than the current semimajor axis of Io. The initial 
ratio of mean motions ($n_1/n_2$) was then 2.234. Since the satellites were 
placed above the synchronous orbit ($n_1 < \Omega$), the effect of the tidal 
friction was to increase the orbit of Io. During the whole experiment, $a_1$ 
increases, after capture into resonance, from $5.9 R_0$ to $9.8 R_0$. 

\begin{figure}[th!]
\figeps{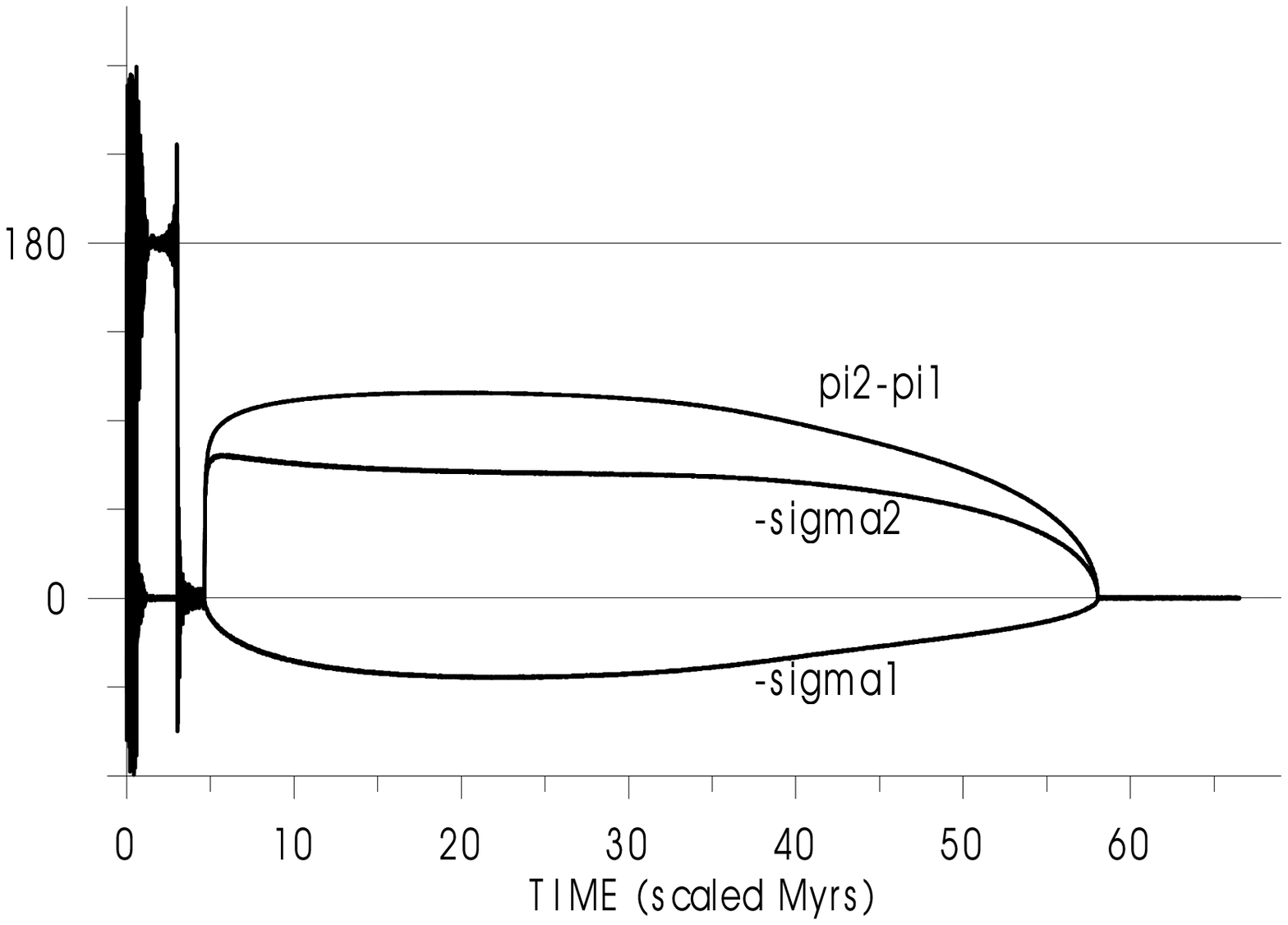}{45mm}
\figeps{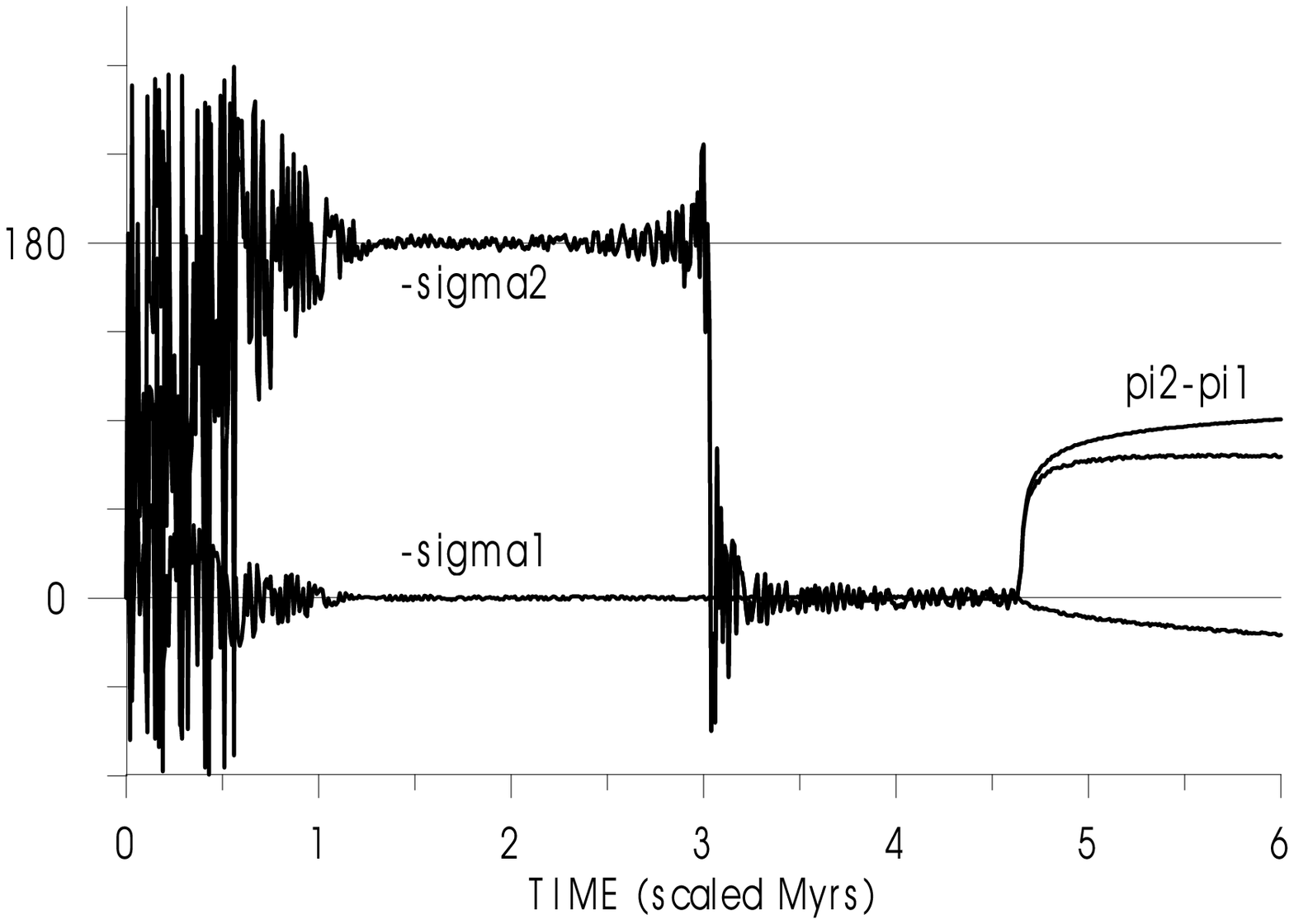}{45mm}
\caption{{\it Top}: Evolution of the angles 
$\Delta\varpi=(\varpi_2-\varpi_1)$, 
$-\sigma_1=\lambda_1-2\lambda_2+\varpi_1$ and 
$-\sigma_2=\lambda_1-2\lambda_2+\varpi_2$ after trapping in the 2/1 resonance.
{\it Bottom:} Enlargement of the initial part.}
\label{fig7}
\end{figure}

The system evolves with the innermost satellite receding from the planet up to 
the moment where the system is captured into a resonance. In the example 
presented in this section, the 2/1 resonance is soon reached ($t \sim 1.2$ Myr 
in the scaled time) and the system is trapped in this resonance. 
Figure \ref{fig5}, shows the evolution of the eccentricities after capture. It 
is worth noting that the variation is not nearly monotonic as seen in 
simplified models. The values increase and decrease following the position of 
the perihelia of the orbits (figure \ref{fig7}), a coupling that is usually 
not considered in many studies. The most notorious phenomenon is the elbow 
discontinuities at $t \sim 4.65$ and $t \sim 58$ scaled Myrs. The first elbow 
occurs when the two orbits cease having parallel semimajor axes and the second 
elbow corresponds to a return to parallelism after a long evolution through 
stationary solutions with non-aligned apses. In another event, which occurs 
at $t \sim 3.05$ scaled Myrs, the eccentricity of the inner satellite becomes 
close to zero and the alignment of the semimajor axes change from 
anti-parallel to parallel. (From $\Delta\varpi = 180^\circ$ to 
$\Delta\varpi = 0^\circ$.) The several phases of the evolution in low 
eccentricities may be compared to the phase portraits of 2/1-resonant 
planetary systems, given by Callegari {\it et al.} (2003). The mode III of 
Callegari {\it et al} has a similar behavior showing stationary solutions 
changing from $\Delta\varpi = 180^\circ$ to $\Delta\varpi = 0^\circ$ when a 
given energy level is crossed.

It is worth noting that this behavior is not reported in the literature, even 
in complex models such as those studied by Gomes (1998) and 
Murray {\it et al.} (2002). We can only assume that their simulations were not 
extended for a long enough time to allow these late stages of the evolution 
of the system to appear.

Comparing these results with the observed configuration of real bodies, we
note that the current orbits of Io and Europa have anti-aligned periapses as 
those of the given example before $t \sim 3.05$. We stress, however, that the 
time scale of our experiment cannot be compared with that of the real 
satellite system, for several reasons. The most important is that our 
simulations started at a time when the bodies were already at the brink of the 
resonance. Of a lesser importance, but significant enough to keep in mind if 
the real Io-Europa pair is meant, is that the adopted time scale flows at 
least 400 times faster than the real one. Additionally, in the actual Galilean 
system, the Laplacian resonance has prevented the Io-Europa subsystem from 
reaching the deeper libration zone of the 2/1 resonance. Thus, it is unlikely 
that these satellites will ever display such turnabouts in libration, even in 
the far future. For extrasolar planetary systems, the current orbits of the 
two planets orbiting the star Gliese 876 are close to have aligned periapses 
as shown in the above simulations between $t\sim 3.05$ and $t\sim 4.65$. 
According to Lee and Peale (2002, 2003), this system may have evolved during 
a certain time in interaction with a remnant dust disk, but the evolution 
stopped when the disk dissipated. As shown by Beaug\'e {\it et al.} (2003), 
the actual orbital parameters of this system are somewhat different of those 
corresponding to an exact resonant stationary solution with aligned periapses.
We may expect that the accumulation of precise observations may give slightly 
different orbital elements or, a second possibility, confirm that the 
periapses are not actually aligned but oscillating about the exact alignment.

\begin{figure}[ht!]
\figeps{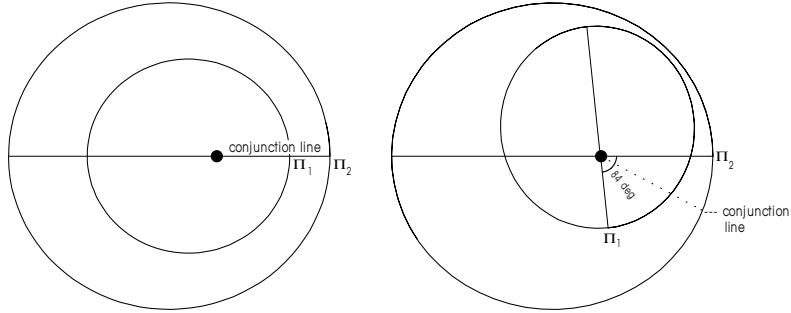}{45mm}
\caption{{\it Left:} Symmetric stationary solution with aligned periapses.
{\it Right:} Asymmetric stationary solution with $|\Delta\varpi| = 84^\circ$. 
The periapses are the points $\Pi_1$ and $\Pi_2$.
The eccentricities are $ e_1=0.286$ and $e_2=0.3$.}
\label{fig13}
\end{figure}

From the dynamical point of view, the more striking behavior is that appearing 
in the interval between $t\sim 4.65$ and $t\sim 58$ scaled Myrs. During this 
time, the system passes through a sequence of stationary solutions in which 
the two periapses are fixed one with respect to another. However, 
$\Delta\varpi$ is no longer $0^o$ or $180^o$ as before. In the beginning of 
this interval, $\Delta\varpi$ increases very fast from $0^o$ to about $80^o$, 
then, it continues increasing up to $100-110^o$ and, thereafter, decreases 
slowly to zero again. These stationary solutions show asymmetric librations 
of the angles $\sigma_1$ and $\sigma_2$, a phenomenon until recently only 
known in the restricted asteroidal case (Beaug\'e 1994, Jancart {\it et al.} 
2002). For illustration purposes, figure \ref{fig13} shows, on the right side, 
planetary orbits in the case of an asymmetric stationary solution where the 
periapses $\Pi_1$ and $\Pi_2$ are separated by $84^o$. That figure also shows, 
on the left side, a symmetric stationary solution with the same eccentricities 
(no matter if not stable). In the {\underline {symmetric}} stationary 
solution with aligned periapses, we have $\sigma_1 = \sigma_2 = 0$, meaning 
that, not only the two periapses are aligned, but the planets have symmetric 
pericentric conjunctions in which both masses pass by the periapses, 
simultaneously, once at each synodic period. The distance of the planets, at 
the symmetric conjunctions, is $a_2(1-e_2)-a_1(1-e_1)$, a value that decreases 
rapidly as $e_2$ grows. In the {\underline{asymmetric}} stationary solutions, 
the periapses shift away one from another and the conjunctions take place at 
an intermediate position. This asymmetry allows the quantity 
$a_2(1-e_2)-a_1(1-e_1)$ to approach zero, and even change sign, without 
necessarily leading to an actual collision. Nevertheless, since conjunction 
may occur at places where both planets come very close to each other, we 
cannot expect that the solutions continue to be stable for very large masses.
Beaug\'e {\it et al.} (2003) have shown that these orbits may only exist for 
planet masses less than $\sim 10^{-2}$ of the central body.

\begin{figure}[hb!]
\figeps{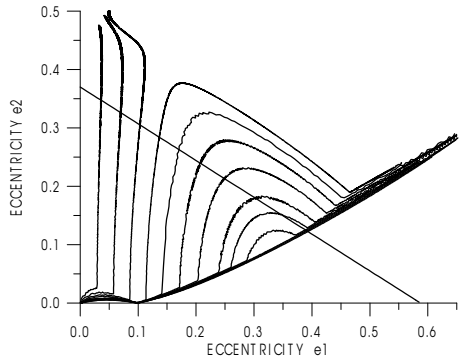}{5cm}
\caption{Families of stationary solutions of the 2/1 resonance shown in the
plane $(e_1,e_2)$. Inner mass is $\,m_1=4.5\times 10^{-5}$. Each curve 
represents a different value of $m_2$; From left to right 
$m_2= 0.5, 1.0, 1.5, 2.0, 2.5, 3.0, 3.5, 4.0, 4.25, 4.5 \times 10^{-5}$. The 
lines corresponding to $m_2 = 4.75, 5.0, 5.25 \times 10^{-5}$ accumulate at 
the bottom of the set. This accumulation marks the separation between the 
region of asymmetric stationary solutions (above) and symmetrical ones (on and 
below). The diagonal straight line across the figure indicates the limit above 
which the two orbits necessarily intersect.}
\label{fig14}
\end{figure}

\begin{figure}[h!]
\figeps{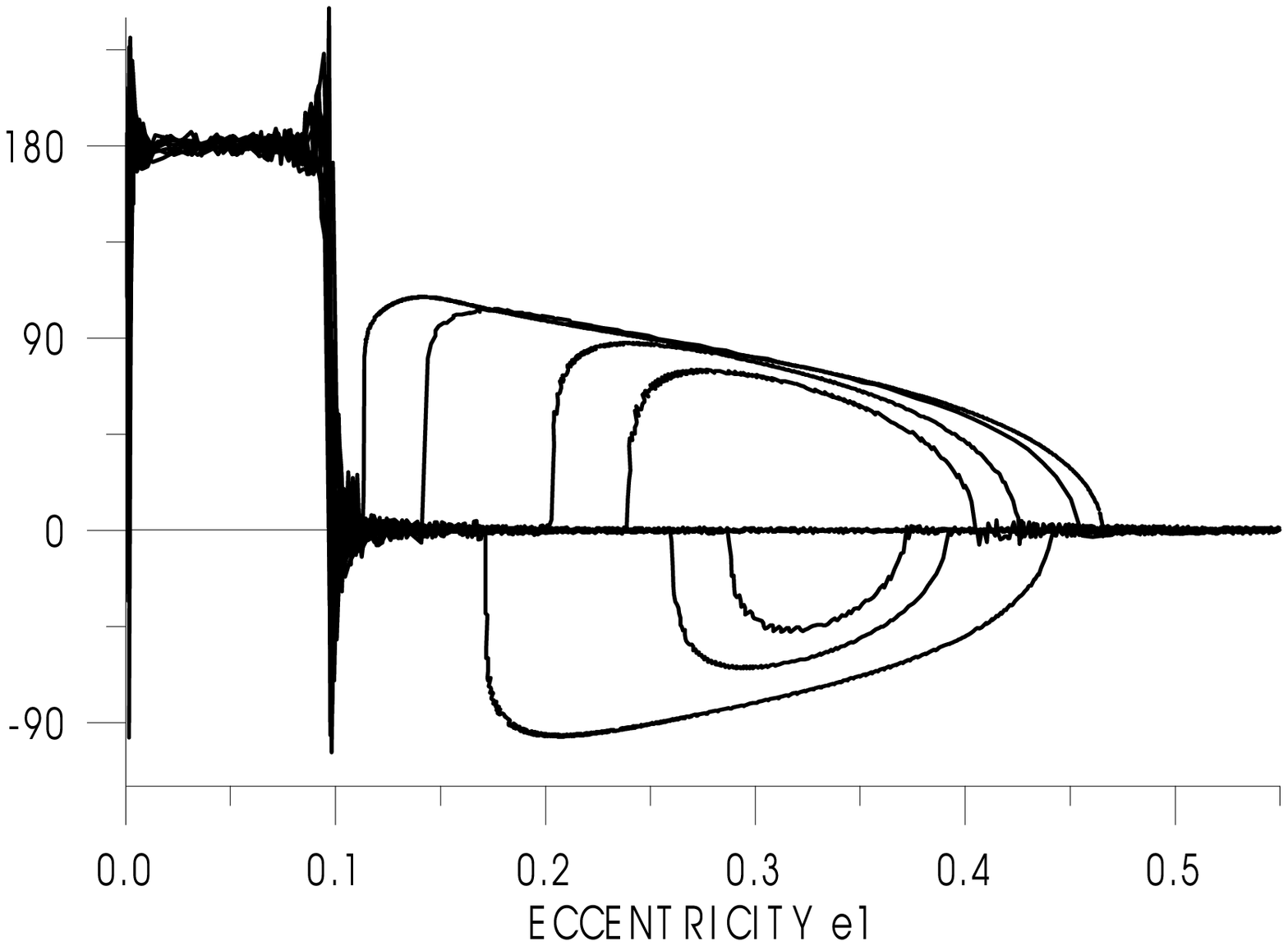}{45mm}
\figeps{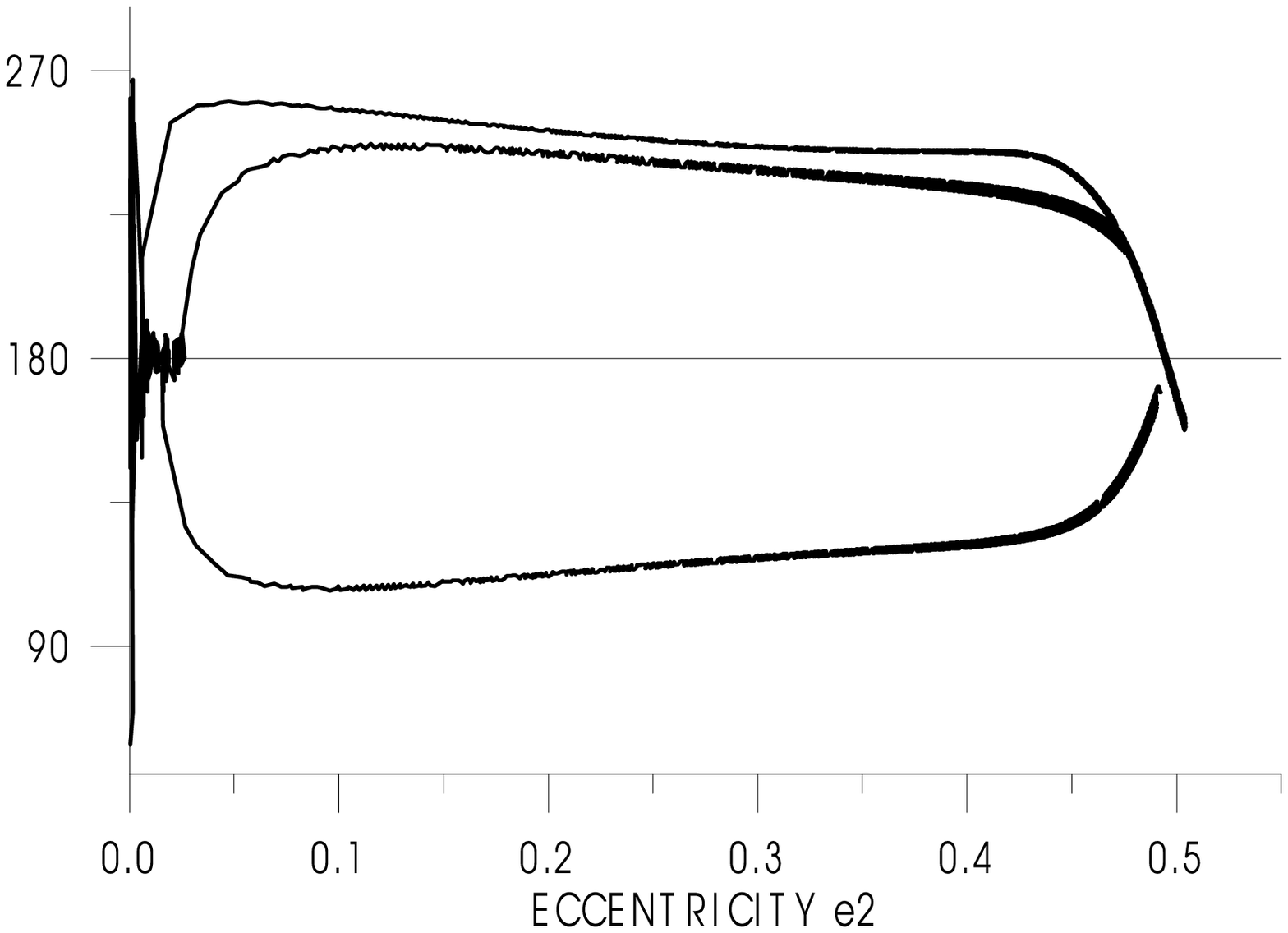}{45mm}
\caption{Values of $\Delta\varpi$ for the families of stationary solutions 
shown in figure \ref{fig14}. 
{\it Top:} $m_2 \ge 2.0 \times 10^{-5}$: the asymmetric solutions bifurcate 
from $\Delta\varpi=180^o$. For $m_2 \ge 5.0 \times 10^{-5}$ no asymmetry 
appears and $\Delta\varpi$ does not bifurcate. 
{\it Bottom:} $m_2 \le 1.5 \times 10^{-5}$: the asymmetric solutions bifurcate 
from $\Delta\varpi=180^o$.}
\label{fig15-16}
\end{figure}

\section{Asymmetric Stationary Solutions}

In the restricted asteroidal problem, it is known that asymmetric solutions 
only exist in the exterior case, that is, when the asteroid is moving in an 
orbit exterior to the planet orbit. Even then, these solutions are only 
detected in resonances of the type $p/1$, being 2/1 and 3/1 the most important 
ones (Beaug\'e 1994). In this section, we search for analogous asymmetric 
solutions in the case of two finite masses, and study their dependence with 
$m_1/m_2$. In particular, we are interested in detecting the minimum value of 
the mass ratio for which these solutions are still present. According to 
Hadjidemetriou (2002), symmetric periodic orbits in the planetary 2/1 
resonance are only stable for values of $m_1/m_2 < 1$. We will see that 
asymmetric solutions allow for a larger range of masses.

In these experiments, $k_2\Delta t$ was increased to 15 s, a value 5 times 
larger than that used in the experiment discussed in previous section. This 
was a limit choice for this parameter; indeed, we can see small fluctuations 
in some of the lines indicating that the variations were not adiabatic enough 
to keep the solutions stationary, forcing small oscillations about the 
stationary solution.

\subsection{2/1 Resonance}

For this commensurability, the mass $m_1$ of the inner orbiting body was fixed 
at $4.5\times 10^{-5}$ and the outer mass $m_2$ was varied in the range 
$0.5 - 5.25\times 10^{-5}$. The results are shown in figures \ref{fig14} and 
\ref{fig15-16}. 

Figure \ref{fig14} shows that asymmetric stationary solutions exist for all 
values of $m_1/m_2$ above a limit close to 1 ($m_1/m_2 > \sim 0.97$). As the 
ratio $m_1/m_2$ decreases approaching this limit, the interval of 
eccentricities where asymmetric solutions exist decreases to zero. For almost 
all mass-ratios within this range, asymmetric stationary solutions in which 
the two orbits cross one another exist. The inclined straight line in figure 
\ref{fig14} shows the values of $e_1$ and $e_2$ such that the apocentric 
distance of $m_1$ equals the pericentric distance of $m_2$. For all initial 
conditions above this curve, the two orbits may intersect.

The equilibrium values of $\Delta\varpi$ are shown in figure \ref{fig15-16}. 
In the evolution paths with $m_2 \ge 2.0\times 10^{-5}$, the asymmetric 
solutions bifurcate from symmetric solutions with aligned periapses; at 
variance, for $m_2 \le 1.5\times 10^{-5}$, the asymmetric solutions bifurcate 
from low-eccentricity symmetric solutions with anti-aligned periapses. The 
two situations are shown in separate figures also because, for 
$m_2 \le 1.5\times 10^{-5}$, the eccentricity $e_1$ has only a very small 
variation, while the $e_2$ has a large variation and thus the variations are 
more clear shown if it lies in the horizontal axis.

It is worth comparing these results on the 2/1 resonance with the periodic 
orbits of Hadjidemetriou (2002). His conclusion that stable solutions are only 
found with $m_1/m_2 < 1$ is true only if we restrict the domain to large 
eccentricity symmetric orbits. In the general case, stable stationary 
solutions can be found for much higher values of the mass ratio.

One last point to mention is that the asymmetric solutions appear as a 
bifurcation point where symmetric solutions change from stable to unstable. 
This bifurcation generates two distinct families of asymmetric solutions, 
each independent of the other. When a trapping occurs, the system may be 
captured in either one or the other of the two new centers. This is clearly 
seen in figures \ref{fig15-16} where approximately half of the solutions 
departed in one direction and half in the opposite one.

\subsection{3/1 Resonance}

\begin{figure}[ht!]
\figeps{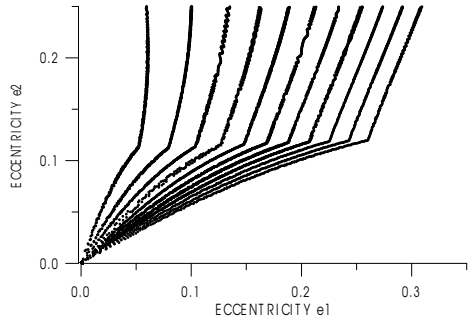}{45mm}
\caption{Families of stationary solutions of the 3/1 resonance shown in the 
plane $e_1,e_2$; $\,m_1=4.5\times 10^{-5}$; from left to right 
$m_2= 0.5, 1.0, 1,5, 2.0, 2.5, 3.0, 3.5, 4.0, 4.5, 5.0, 5.5 \times 10^{-5}$.
The line formed by the elbows of the curves, at $e_2 \sim 0.11$, separates 
the symmetric (below) and asymmetric (above) stationary solutions.}
\label{fig17}
\end{figure}

\begin{figure}[ht!]
\figeps{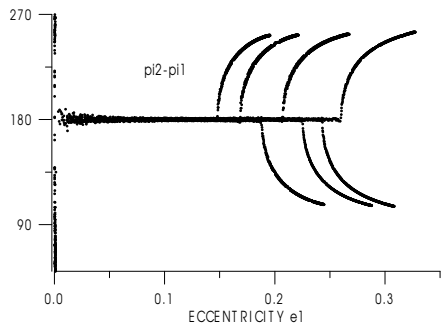}{45mm}
\caption{Values of $\Delta\varpi$ for the families of stationary solutions 
shown in figure \ref{fig17} for which $m_2 \ge 2.5 \times 10^{-5}$; 
In the 3/1 resonance the asymmetric solutions bifurcate from symmetric 
solutions with anti-aligned periapses ($\Delta\varpi=180^o$)}
\label{fig18}
\end{figure}

In the restricted asteroidal problem, asymmetric solutions exist for all 
exterior resonances of the type $p/1$ (i.e. 2/1, 3/1, 4/1, etc.). A series 
of simulations was then done to see whether the general three-body problem 
also presented asymmetric points in the 3/1 resonance. As in the previous 
subsection the relative mass of the innermost orbiting body was fixed at 
$4.5\times 10^{-5}$ and the outer value was varied from $2.5\times 10^{-5}$ 
to $5.5\times 10^{-5}$ in steps of $0.5\times 10^{-5}$. The remaining 
parameters were chosen as:
\begin{displaymath}
\begin{array}{l@{\hspace{2cm}}l}
a_{1(0)} = 5.85 R_{0}	& 	a_{2(0)} = 12.4 R_{0}\\
e_{1(0)} = 0.0010 &	e_{2(0)} = 0.0004 \\
\end{array}\end{displaymath}
The results are shown in figs. \ref{fig17} and \ref{fig18}. The capture into 
resonance occurs quickly and $\Delta\varpi$ reaches $180^o$ remaining there up 
to the bifurcation and switching to an asymmetric stationary solution. Periodic
orbits with aligned periapses were not seen, although, in the very beginning 
of the simulations $\Delta\varpi$ appeared temporarily oscillating about $0^o$ 
with a large amplitude. The more characteristic feature of the 3/1-resonant 
asymmetric stationary solutions is the boundary between symmetric and 
asymmetric regions at $e_2 \sim 0.11$ (elbows line in figure \ref{fig17}). 
In this resonance, asymmetric stationary solutions exist for all values of 
the mass ratio (greater or smaller than unity) up to the limit corresponding 
to a curve whose elbows occurs for $e_1 \sim 1$. For mass ratios larger than 
this, the stationary solution lines will remain below the $e_2 \sim 0.11$ 
limit for all $e_1 < 1$.

\section{Conclusions}

In this paper, we presented a series of numerical simulations of the evolution 
of systems of two massive orbiting bodies after capture into the 2/1 and 3/1 
resonances. Albeit with a change in the timescale of the solutions, the same 
behavior can represent the evolution of a pair of satellites around one planet 
(the data used were those from the Io-Europa pair) or two exoplanets orbiting 
close to a star. The simulations were done using tides in the central body as 
the source of the non-conservative perturbation, but the results do not depend 
on the particular force used, and should happen in similar way with other 
non-conservative forces provided its effects are sufficiently slow to have 
adiabatic variations. Thus, the same behavior is expected to hold in the 
evolution of exoplanets captured into a 2/1 or 3/1 resonance due to (for 
example) interactions with a residual disk of matter or any other 
non-conservative force. For instance, Lee and Peale (2003) have obtained 
results very similar to those discussed in section 3 in a simulation of a 
system of 2 planets in which the outer planet is driven inward by torques 
exerted on it by outside nebular material. 

Finally, we have shown that the evolution of resonant trapped massive bodies 
is not simple. Our results show a surprising richness of solutions: symmetric 
librations around $\Delta \varpi = 0$ and around $\Delta \varpi = \pi$, 
asymmetric solutions with a wide range of values of $\Delta \varpi$, 
turnabouts from one configuration to another during the secular variation of 
semimajor axes, etc. Many of these solutions exist for large values of the 
eccentricities. As our knowledge of extrasolar planets continues to grow, it 
will be interesting to see how many of these possible configurations actually 
occur in the real world.

\subsection*{Acknowledgements} 
{The authors acknowledge the support of FAPESP and CNPq to this investigation 
and the Instituto de Pesquisas Espaciais, INPE, where C.Beaug\'e was visiting 
investigator during the realization of this investigation. The research 
subject ``Interplay of Tides and Resonances'' was suggested to SFM by Iwan 
Williams.}



\begin{thebibliography}{99}

\bibitem{}
Aksnes, K. and Franklin, F.A.: 2001, ``Secular Acceleration of Io derived from 
Mutual Satellite Events'', {\it Astron. J.} {\bf 122}, 2734-2739. 

\bibitem{}
Beaug\'e, C.: 1994, ``Asymmetric Librations in Exterior Resonances'', {\it Cel.
Mech. Dyn. Astron.} {\bf 60}, 225-248. 

\bibitem{}
Beaug\'e, C., Ferraz-Mello, S. and Michtchenko, T.A.:  2003, ``Extrasolar
Planets in Mean-Morion Resonance: Apsidal and Asymmetric Stationary 
Solutions'', {\it Astroph. J.}, submitted.

\bibitem{}
Callegari, Jr., N., Michtchenko, T.A. and Ferraz-Mello, S.: 2003, ``Dynamics 
of two planets in the 2:1 and 3:2 mean-motion resonances'' (in preparation). 

\bibitem{}
Ferraz-Mello, S.: 1979, {\it Dynamics of the Galilean Satellites}, IAG-USP,
S\~ao Paulo. 

\bibitem{}
Ferraz-Mello, S.: 1987, ``Averaging the Elliptic Asteroidal Problem Near a 
First-Order Resonance''. {\it Astron. J.} {\bf 94}, 208-212. 

\bibitem{}
Gomes, R. S.: 1998: ``Orbital Evolution in Resonance Lock. II. Two Mutually
Perturbing Bodies'', {\it Astron. J.} {\bf 116}, 997-1005. 

\bibitem{}
Hadjidemetriou, J.D.: 2002, ``Resonant Periodic Motion and the Stability of
the Extrasolar Planetary Systems'' {\it Celest. Mech. Dyn. Astron.} {\bf 83},
141-154.

\bibitem{}
Jancart, S.: 2002, {\it R\'esonances et Dissipations}, Dr. Thesis, Presses
Universitaires, Namur.

\bibitem{}
Jancart, S., Lemaitre, A., and Istace, A.: 2002, ``Second Fundamental Model of
Resonance with Asymmetric Equilibria'' {\it Celest. Mech. Dyn. Astron.} 
{\bf 84}, 197-221. 

\bibitem{}
Lee, M.H. and Peale, S.J.: 2002, ``Dynamics and Origin of the 2:1 Orbital
Resonances of the GJ 876 Planets'' {\it Astroph. J.} {\bf 567}, 596-609. 

\bibitem{}
Lee, M.H. and Peale, S.J.: 2003, ``Extrasolar Planets and Mean-Motion
Resonances'' {\it ASP Conf. Series} (in press). 

\bibitem{}
Lieske, J.H.: 1987, ``Galilean Satellites Evolution - 
Observational Evidence for Secular Changes in Mean-Motions'' 
{\it Astron. Astrophys.} {\bf 176}, 146-158.

\bibitem{}
Lieske, J.H.: 1998, ``Galilean Satellites Ephemerides E5'', {\it Astron.
Astrophys. Supp. Ser.} {\bf 129}, 205-217.  

\bibitem{}
MacDonald, G.J.F.: 1964, ``Tidal Friction'' {\it  Rev.Geophys.} {\bf 2},
467-541.

\bibitem{}
Mignard, F.: 1981, ``The Evolution of the Lunar Orbit Revisited. I'', {\it Moon
and Planets}, {\bf 20}, 301-315. 

\bibitem{}
Murray, N., Paskowitz, M. and Holman, M.: 2002, ``Eccentricity Evolution of
Resonant Migrating Planets'', {\it Astroph. J.} {\bf 565}, 608-620.  

\bibitem{}
Peale, S.J.: 1999, ``Origin and Evolution of the Natural Satellites'' 
{\it Ann. Rev. Astron. Astroph.} {\bf 37}, 533-602. 

\bibitem{}
Tisserand, F.: 1896, {\it Trait\'e de M\'ecanique C\'eleste}, Gauthier-Villars,
Paris, vol. IV. 

\bibitem{}
Touma, J. and Wisdom, J.: 1994, ``Evolution of the Earth-Moon System'', 
{\it Astron. J.} {\bf 108}, 1943-1961. 

\bibitem{}
Wu, Y. and Goldreich, P.: 2002, ``Tidal Evolution of the Planetary System
around HD 83443'', {\it Astroph. J.} {\bf 564}, 1024-1027. 

\bibitem{}
Yoder, C.F.: 1979, ``How tidal heating in Io drives the Galilean Orbital
Resonance Locks'',{\it  Nature} {\bf 279}, 767-770. 

\end{thebibliography}
\end{document}